\documentclass[twocolumn,superscriptaddress,showpacs,amsmath,amssymb,floatfix,longbibliography]{revtex4-1}
\usepackage{graphicx}
\usepackage{color}
\usepackage{enumerate}

\newcommand{\la}{\langle}
\newcommand{\ra}{\rangle}

\begin{document}
\title{Geometric allocation approach to accelerating directed worm algorithm}
\author{Hidemaro Suwa}
\affiliation{Department of Physics and Astronomy, University of Tennessee, Knoxville, Tennessee 37996, USA}
\affiliation{Department of Physics, University of Tokyo, Tokyo 113-0033, Japan}

\date{\today}
\begin{abstract}
  The worm algorithm is a versatile technique in the Markov chain Monte Carlo method for both classical and quantum systems. The algorithm substantially alleviates critical slowing down and reduces the dynamic critical exponents of various classical systems. It is crucial to improve the algorithm and push the boundary of the Monte Carlo method for physical systems. We here propose a directed worm algorithm that significantly improves computational efficiency. We use the geometric allocation approach to optimize the worm scattering process: worm backscattering is averted, and forward scattering is favored. Our approach successfully enhances the diffusivity of the worm head (kink), which is evident in the probability distribution of the relative position of the two kinks. Performance improvement is demonstrated for the Ising model at the critical temperature by measurement of exponential autocorrelation times and asymptotic variances. The present worm update is approximately 25 times as efficient as the conventional worm update for the simple cubic lattice model. Surprisingly, our algorithm is even more efficient than the Wolff cluster algorithm, which is one of the best update algorithms. We estimate the dynamic critical exponent of the simple cubic lattice Ising model to be $z \approx 0.27$ in the worm update. The worm and the Wolff algorithms produce different exponents of the integrated autocorrelation time of the magnetic susceptibility estimator but the same exponent of the asymptotic variance. We also discuss how to quantify the computational efficiency of the Markov chain Monte Carlo method. Our approach can be applied to a wide range of physical systems, such as the $| \phi |^4$ model, the Potts model, the O($n$) loop model, and lattice QCD.
\end{abstract}

\maketitle

\section{Introduction}
\label{sec:intro}
The Markov chain Monte Carlo (MCMC) method is a powerful numerical tool for studying a wide variety of statistical mechanical problems\,\cite{LandauB2005,NewmanB1999}. Many kinds of non-trivial phases and phase transitions in both classical and quantum systems have been uncovered by the MCMC method. The essence of the method is to construct a global transition kernel as a series of {\em local} kernels acting on local state variables. One can sample states from an arbitrary target distribution even in a vast number of dimensions (or degrees of freedom) of state space.

In the MCMC method, one has to care about autocorrelation between samples. The autocorrelation function\,\cite{LandauB2005,NewmanB1999} of an estimator $\hat{\mathcal O}$ is defined by
\begin{equation}
A_{\hat{\mathcal O}} (t) = \frac{\langle {\mathcal O}_{i+t}{\mathcal O}_i \rangle - \langle \hat{\mathcal O} \rangle^2}{\langle \hat{\mathcal O}^2
  \rangle - \langle \hat{\mathcal O} \rangle^2},
\label{A}
\end{equation}
where ${\mathcal O}_s$ is the sample of a physical quantity ${\mathcal O}$, such as the total energy, at the $s$-th Monte Carlo step. The Monte Carlo average is denoted by the bracket $\langle \cdot \rangle$. The autocorrelation function eventually becomes (almost) independent of $i$ in Eq.~(\ref{A}) after the distribution convergence, namely the thermalization (the burn-in). In many cases, the function decays exponentially for large $t$:
$
  A_{\hat{\mathcal O}}(t) \sim e^{ - t / \tau_{{\rm exp}, {\hat{\mathcal O}}}} \label{Ad},
$
where
\begin{equation}
  \tau_{{\rm exp},{{\hat{\mathcal O}}}} = \limsup_{t \to \infty} \frac{ t }{ - \ln | A_{\hat{\mathcal O}}(t)| }
\end{equation}
is the exponential autocorrelation time of $\hat{\mathcal O}$. Autocorrelation reduces the effective number of independent Monte Carlo
samples to $M_{\rm eff} \approx M / 2 \tau_{{\rm int},{\hat{\mathcal
      O}}}$, where $M$ is the number of samples obtained in a simulation and
\begin{equation}
  \tau_{{\rm int}, {\hat{\mathcal O}}}= \frac{1}{2} + \sum_{t=1}^{\infty} A_{\hat{\mathcal O}} (t) \label{tau_int}
  \end{equation}
is the integrated autocorrelation time of $\hat{\mathcal O}$. The constant $\frac{1}{2}$
comes from the discrete nature of the Monte Carlo time evolution. The needed
computation time for a certain precision is proportional to these autocorrelation times, $\tau_{{\rm exp}, {\hat{\mathcal O}}}$ and $\tau_{{\rm int},
  {\hat{\mathcal O}}}$. They may depend on estimators and update
methods.

The MCMC method can be applied to many kinds of phase transitions in principle, but the convergence (relaxation) rate and the sampling efficiency can become very poor in some cases, such as critical slowing down\,\cite{HohenbergH1977,Sokal1997}. As the system approaches a critical point, the exponential autocorrelation time diverges: $\tau_{\rm exp} \propto \xi^z \propto |t|^{-\nu z}$, where $\xi$ is the correlation length, $t$ is the temperature difference from a critical point, and $\nu$ is the critical exponent of the correlation length. The exponent $z$ is called the dynamic critical exponent, which is given by
\begin{equation}
  z = \lim_{L \to \infty} \frac{ \displaystyle \ln \left( \max_{\hat{\mathcal O}} \tau_{{\rm exp}, {\hat{\mathcal O}}} \right)}{\ln L}
\end{equation}
at the critical point, where $L$ is the system length. Note that most estimators share the maximum exponential autocorrelation time. Thus, $ \tau_{{\rm exp}, {\hat{\mathcal O}}} \propto L^z$ asymptotically at the critical point. The exponent of the integrated autocorrelation time may differ from $z$, but they are identical in many cases. For example, in the case of the square lattice Ising model, the Metropolis algorithm for the single spin update suffers from the rapid growth of the autocorrelation times: $ \tau_{\rm exp} \sim \tau_{\rm int} \propto L^z$ with $z \approx 2.17$\,\cite{WangH1997,NightingaleB2000,MuraseI2008,LiuPS2014}. The dynamic critical exponent is expected to be universal among many MCMC updates\,\cite{NightingaleB2000}. Such a large dynamic critical exponent hampers efficient sampling near a phase transition: the spectral gap of a Markov chain $\Delta \approx \tau_{\rm exp}^{-1} \propto \xi^{-z}$ is reduced to zero at the critical point.  It is thus crucial to devise a smart update method that alleviates or avoids slowing down.

In the case of unfrustrated models, the cluster algorithms, such as the Swendsen-Wang\,\cite{SwendsenW1987} and the Wolff\,\cite{Wolff1989} algorithms, reduce the dynamic critical exponent significantly\,\cite{TamayoBK1990,CoddingtonB1992,LiuPS2014}: for example, $z \approx 0.3$ for the Ising model in two dimensions. The Wolff algorithm is known to be more efficient than the Swendsen-Wang algorithm in $d \geq 3$ dimensions. The size of a cluster corresponds to the correlation length, and the flip of clusters, which can be performed with probability one, achieves an efficient non-local spin update. Forming such an efficient cluster, however, is non-trivial or impractical in general cases. The application of the cluster updates is thus limited to specific models.

In the meantime, the worm algorithm has been one of the most versatile techniques in the worldline quantum Monte Carlo method\,\cite{ProkofievST1998,SyljuasenS2002}. In quantum cases, a naive local update of worldlines is often not allowed: for example, a local spin flip breaking up the worldline is not allowed in the $XXZ$ quantum spin model because the total magnetization is conserved by the Hamiltonian. The worm algorithm works especially well for cases in which the allowed configurations are restricted by such constraints.

The main idea of the worm algorithm is to achieve an eventual non-local update resulting from sequential local updates in extended state space. In practice, the extended space is composed of configurations that contain kinks, which break the constraint. We insert a pair of kinks, which is called the worm, and move one of them, which is called the worm head.

The whole procedure of the worm algorithm is described by the repetition of the following processes: (i) A pair of kinks is inserted at a randomly chosen position of the system. (ii) One of the kinks moves in a stochastic way, updating the configuration. (iii) When meeting each other, the kinks are removed. 

The worm algorithm for classical systems\,\cite{ProkofievS2001} was proposed as well, which we call the classical algorithm hereafter. During the process (ii) mentioned above, the position of the worm (the kinks) randomly shifts from site to site of a lattice. The algorithm aims at a random walk of the kink at sites (vertices). The next site is chosen at random among the nearest sites. The worm shifting process is then accepted or rejected using the Metropolis algorithm. The detailed balance holds in every shifting process. Although each worm move is local in the extended space, a non-local update in the original space is eventually achieved after the whole worm update (from insertion to removal). Despite its local nature, the worm algorithm significantly reduces the dynamic critical exponents for several classical models\,\cite{DengGS2007,LiuDG2011}. We review the detail of the classical algorithm for the Ising model in Sec.~\ref{alg1}.

It is critical to optimize the stochastic worm update for efficient computation. How can we improve the worm algorithm? The stochastic worm move can be viewed as a diffusion process of the kink in the real space. Thus, higher diffusivity of the kink is expected to yield higher sampling efficiency. In particular, the worm backscattering process, which cancels the previous update, should be averted for efficient sampling.

The directed loop (or the directed worm) algorithm was proposed to improve the efficiency of the worldline quantum Monte Carlo method\,\cite{SyljuasenS2002}. The directed worm has an additional feature, the direction to move in. The update does not hold the detailed balance for each local worm process but does for the whole worm update from insertion to removal. Thanks to the directed path, backscattering is successfully suppressed.

In the meantime, the geometric allocation approach was proposed to optimize the transition probability in a flexible manner\,\cite{SuwaT2010}. It is a versatile technique for the MCMC method. The basic concept of this approach is that {\em the flows} between the states are purposefully allocated in a geometric fashion. One can easily find a set of probabilities that holds the global (total) balance even without detailed balance and minimize the rejection probability. The efficiency of the directed worm update in the worldline quantum Monte Carlo method is significantly improved by the geometric allocation approach\,\cite{SuwaT2010,Suwa2014}.

The purpose of the present paper is to enhance the diffusivity of the kink in the worm algorithm. We propose a directed worm algorithm accelerated by the geometric allocation approach. The key ideas of our approach are the following: 1) The kinks are located on {\em bonds} (edges) instead of sites of a lattice. 2) The worm move is directed. 3) The worm backscattering probability is minimized, and the forward scattering probability is maximized using the geometric allocation. We confirm enhanced diffusivity by calculating the probability distribution of the relative position of the two kinks. The present algorithm is detailed in Sec.~\ref{alg2}.

We also discuss how to compare MCMC samplers in Sec.~\ref{mc}. We stress that the sampling efficiency of the MCMC method should be quantified by the asymptotic variance, the prefactor of the asymptotic scaling of the statistical error squared.

We demonstrate, in Sec.~\ref{result}, that the present worm algorithm for the Ising model significantly improves computational efficiency. We show that the efficiency of the present worm update is approximately 25 times as high as that of the classical worm update for the simple cubic lattice Ising model at the critical temperature. There is no extra computational cost in the present algorithm, as compared to the classical algorithm. Our algorithm is even more efficient than the Wolff cluster algorithm, which is one of the best update methods for the Ising model. We estimate the dynamic critical exponent of the simple cubic lattice Ising model to be $z \approx 0.27$ in the worm update.

Our approach is applicable to many physical models, such as the $| \phi|^4$ model\,\cite{ProkofievS2001}, the Potts model\,\cite{MercadoEG2012}, the O($n$) loop model\,\cite{JankeNS2010,LiuDG2011,ShimadaJK2014}, and lattice QCD\,\cite{AdamsC2003}, and expected to improve the efficiency of the MCMC update for these models as well as for the Ising model.

The present paper is summarized with discussions in Sec.~\ref{sd}.

\section{Classical algorithm}
\label{alg1}
We review the conventional (classical) worm
algorithm\,\cite{ProkofievS2001} for the Ising model in this
section. Let the model be represented by $-H/T=K\sum_{\la ij \ra}
\sigma_i \sigma_j$, where $H$ is the Hamiltonian, $T$ is the temperature, and $\sigma_i = \pm 1$ is the Ising spin variable at each site
(vertex) $i$ of a lattice (graph). The partition function of the
canonical ensemble can be represented by
\begin{align}
  Z&=\sum_{\sigma_i = \pm 1} e^{K \sum_{\la ij \ra} \sigma_i \sigma_j}=\sum_{\sigma_i = \pm 1} \prod_{b=\la ij \ra} e^{K \sigma_i \sigma_j} \nonumber \\
  &= \sum_{\sigma_i= \pm 1} \prod_{b=\la ij \ra}  \cosh(K) \sum_{n_b=0,1} [\sigma_i \sigma_j \tanh K]^{n_b} \nonumber \\
  &= 2^{N} [\cosh K]^{N_b^{tot}} \sum_{\{n_b\}}^{loops} [\tanh K]^{\ell} ,
  \label{Z}
\end{align}
where the bond variable on bond $b$ is denoted by $n_b$, the identity $e^{K \sigma_i \sigma_j}= \cosh(K) \sum_{n_b=0,1}[ \sigma_i \sigma_j \tanh K]^{n_b}$ is used in the second line, and $N$ and $N_b^{tot}$ are the total number of sites and bonds of a lattice, respectively. In the last line, the sum runs over all the bond configurations that only have closed loops formed by the activated bonds ($n_b=1$). The configurations that have open strings of activated bonds do not contribute to the partition function. The total length of the closed loops is denoted by $\ell \equiv \sum_{b} n_b$. The bond variables are sampled by means of the MCMC method under the constraint of the loop structure: the number of activated bonds meeting at each site is even. Any set of bond variables can be used as the initial state in the simulation as long as the loop constraint is satisfied. As the initial state, we chose the vacuum state, in which the bond variables are all deactivated ($n_b=0$ $\forall b$).

The worm algorithm is an efficient update method for sampling under such a constraint or a conservation law. The fundamental idea is to extend the state space and allow configurations containing kinks, which break the constraint. Let us consider inserting two kinks and move one of them in a stochastic way. The moving kink is called the worm head, and the other is the worm tail. The classical worm algorithm\,\cite{ProkofievS2001} is described as follows:
\begin{enumerate}[Step 1:]
  \item Choose a site $i_0$ at random as the starting point and set $i \leftarrow i_0$. Insert the worm head and tail at $i_0$. Go to step 2.
  \item Choose a site $j$ at random among the nearest
    neighbor sites of site $i$ and shift the worm head from $i$
    to $j$ with probability $p=[\tanh K]^{1-n_b}$, where $n_b$ ($=0$ or $1$) is
    the bond variable on $b=\la i j \ra$ before the shift. If the shift
    is accepted, update $n_b$ ($0 \leftrightarrow 1$) and set $i
    \leftarrow j$.  If $j = i_0$, go to step 3. Otherwise, repeat step
    2.
    \item Measure observables. Go to step 1 after removing the worm with probability $p_{\rm move}$, or go to step 2 with probability $1-p_{\rm move}$.
\end{enumerate}
The probability $p_{\rm move}$ can be set to an arbitrary value in
$(0,1]$: $p_{\rm move}=1/2$ in Ref.\,\citenum{ProkofievS2001}.

In the measurement, the total energy can be measured by the total number of
activated bonds:
\begin{align}
  E &= - \frac{\partial \ln Z}{\partial \beta} \nonumber \\
  &= - N_b^{total} \tanh K - \left( \frac{1}{\tanh K} - \tanh K \right) \la \ell \ra, \label{e}
\end{align}
where $\beta=1/T$ is the inverse temperature. The spin correlation function, $G_{ij} \equiv {\rm tr} [\sigma_i \sigma_j e^{-\beta H}] / Z$, can be estimated by $\la N_{ij} \ra / \la N_j \ra$, where $N_{ij}$ is how many times the head is at site $i$ and the tail is simultaneously at site $j$ in step 2, and $N_j$ is how many times the head and tail are both at site $j$ in step 3. The magnetic susceptibility, $\chi \equiv \frac{\beta}{N}\sum_{ij}G_{ij}$, can be estimated by
\begin{equation}
  \chi = \beta \la \ell_{\rm worm} \ra,
  \label{chi-cl}
\end{equation}
where $\ell_{\rm worm}$ is the worm length, the total number of worm shifting processes in step 2 including the rejection process. It is straightforward to calculate the Fourier transformed correlation function: one only needs to take into account a phase factor depending on the kink position. The correlation length can be calculated using the Fourier transformed correlation functions and the moment method\,\cite{SuwaT2015}.

The worm algorithm significantly reduces the dynamic critical exponents of several models\,\cite{DengGS2007}. It has been applied to many fundamental physical systems, such as the Potts model\,\cite{MercadoEG2012}, the $| \phi |^4$ model\,\cite{ProkofievS2001}, the O($n$) loop model\,\cite{JankeNS2010,LiuDG2011,ShimadaJK2014}, and lattice QCD\,\cite{AdamsC2003}.

One can use the worm algorithm for dual variables on a dual lattice\,\cite{ProkofievS2001,HitchcockSA2004,Wang2005,RakalaD2017}. The dual worm algorithm samples domain walls of the original spin variables; in other words, it samples ``unsatisfied'' bonds that increase the total energy. While the classical worm algorithm is formulated in the high temperature expansion, the dual worm algorithm is in the low temperature expansion for the dual inverse temperature: $\beta' = - \frac{1}{2} \ln \tanh \beta$\,\cite{KramersW1941,Kogut1979}. One of the advantages of the dual worm algorithm is that it is applicable also to frustrated cases, in which the original worm algorithm suffers from the negative sign problem\,\cite{Wang2005,RakalaD2017}.

In two dimensions, the dual variables are under an additional constraint: the winding number of the loops formed by the unsatisfied bonds is even (odd) for periodic (antiperiodic) boundaries. If this constraint is ignored in simulation, nevertheless, the domain wall free energy, the difference between the free energy of the system with periodic boundaries and the free energy of the system with antiperiodic boundaries, can be estimated from the winding number histogram\,\cite{HitchcockSA2004}: $e^{-\beta ( F_{\rm AP} - F_{\rm P} )} = Z_{\rm AP} / Z_{\rm P} = \la N_{\rm AP} \ra / \la N_{\rm P} \ra$, where $F_{\rm AP}$ is the free energy of the system with periodic (antiperiodic) boundaries in one (the other) direction, $F_{\rm P}$ is the free energy of the system with periodic boundaries in both directions, $Z_{\rm AP}$ and $Z_{\rm P}$ are the associated partition functions, $N_{\rm AP}$ is how many times the winding number is even (odd) in one (the other) direction, and $N_{\rm P}$ is how many times the winding numbers in the two directions are both even, respectively. Because the square lattice is self-dual, the dual worm update at the critical temperature is identical to the original worm update, except for the winding number constraint. The domain wall free energy is, therefore, accessible in both formalisms.

\section{Present approach}
\label{alg2}
We present a modified worm update in this section. The worm backscattering (rejection) probability is minimized, and the forward scattering probability is maximized using the geometric allocation approach. Our algorithm is indeed free from rejection at the critical temperatures of the square lattice and the cubic lattice Ising models. As a result, the diffusivity of the worm head is enhanced, which improves computational efficiency. We show the ergodicity of the Markov chain created by the present method and describe how to measure relevant physical quantities, such as the magnetic susceptibility. We also discuss a possible bias and how to avoid it in the worm algorithm.

\subsection{Worm on bonds}
\label{bw}
We adopt the same representation of the partition function [Eq.~(\ref{Z})] with the classical worm algorithm. Our goal is to sample bond variables $\{n_b\}$ efficiently under the loop constraint. We here consider inserting the worm, namely a pair of kinks, on a bond, or an edge, of a lattice. Our worm is distinct in this respect from the classical worm, which is always located at sites. We then move the worm head, that is, one of the pair, from one bond to another in a stochastic way: when coming to a site, the worm head scatters to another (or possibly the same) bond with a certain probability. This scattering process continues until the head comes back to the tail, that is, the other of the pair.

\begin{figure}  
\begin{center}
\includegraphics[width=5cm]{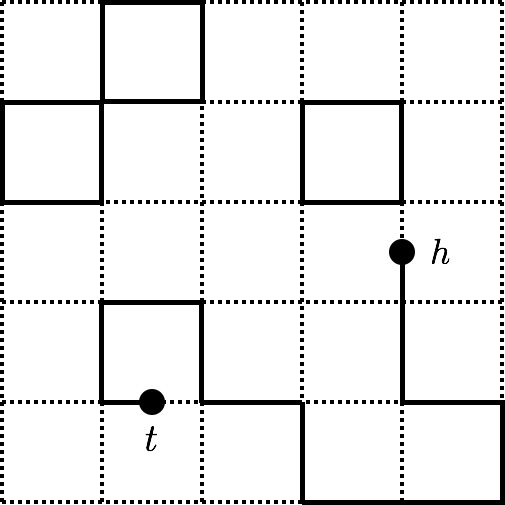}
\caption{Example of a configuration containing the present worm in the square lattice Ising model. The solid lines show the activated bonds, and the broken lines show the deactivated bonds. The solid circles indicate the worm head ($h$) and the worm tail ($t$), both of which break the loop constraint of the activated bonds.}
\label{fig:example}
\end{center}
\end{figure}

A typical configuration containing the present worm in the $L=6$ square lattice Ising model with open boundaries is illustrated in Fig.~\ref{fig:example}. In our algorithm, each kink is located at {\em the center} of a bond; the bond variables can take $n_b=0, \frac{1}{2}$, or $1$. The head has a moving direction in a fashion similar to the directed loop algorithm\,\cite{SyljuasenS2002}.

\begin{figure}
\begin{center}  
\includegraphics[width=8cm]{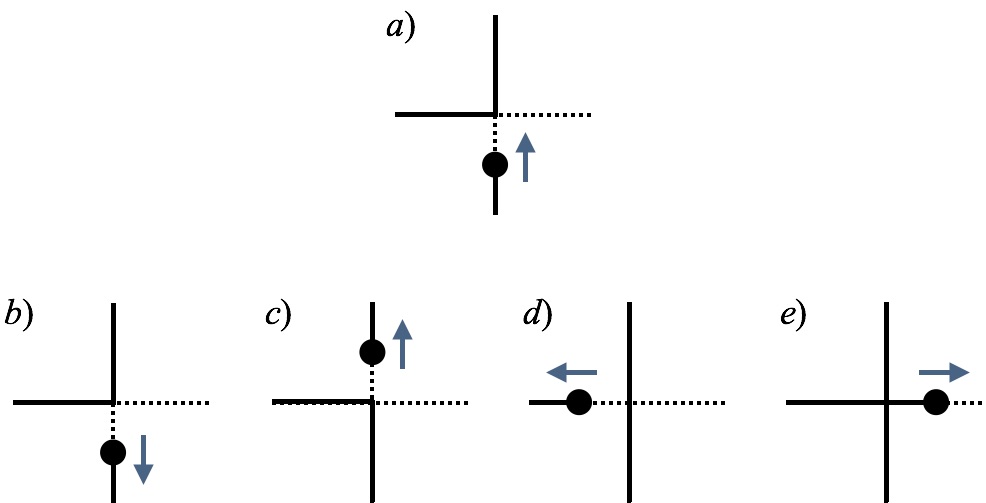}
\caption{Example of the worm scattering process for the square lattice case. The solid circle in each graph shows the worm head, and the arrow shows the moving direction of the head. When coming to a vertex (state $a$), the worm head scatters to another (or possibly the same) bond (states $b$, $c$, $d$, and $e$) with a certain probability.}
\label{fig:ws}
\end{center}
\end{figure}

Suppose the moving direction of the head is upward in Fig.~\ref{fig:example}. Then the head scatters at the next site (vertex) and moves to a bond connecting to the site stochastically, which we call the worm scattering process. The four possible states after the scattering are shown as $b$, $c$, $d$, and $e$ in Fig.~\ref{fig:ws}. The next state is chosen between the four states with a certain probability. We discuss probability optimization in Sec.~\ref{go}. After the worm scattering, bond variables are updated, as shown in Fig.~\ref{fig:ws}; the halves of bonds are updated ($n_b=0,\frac{1}{2},1$) as the kink is assumed to be at the center of a bond. We repeat this worm scattering process until the head comes back to the tail position.

The whole procedure of the present algorithm is described as follows:
\begin{enumerate}[Step 1:]
\item Choose a bond $b_0$ at random as the starting point and $b \leftarrow b_0$. Insert the worm head and tail at the center of $b_0$. Choose the moving direction at random. Go to step 2.
\item Choose the next bond $c$ with the probability optimized using the geometric allocation. If $b \neq c$, update the bond variables $n_b$ and $n_c$, and set $b \leftarrow c$. If $b = b_0$, go to step 3. Otherwise, repeat step 2.
  \item Measure observables and go to step 1 after removing the kinks (worm).
\end{enumerate}
As compared with the classical worm algorithm, the probability $p_{\rm move}$ at step 3 is fixed to one in our algorithm. Note that the present worm carries extra weight (a factor of a half) such that the insertion and the removal are accepted with probability one, as discussed in Sec.~\ref{obs}.

One of the advantages of our approach is that it is straightforward to optimize the worm scattering probability. In general, the transition probability is set under global balance in the MCMC method. If one did not resort to the Metropolis algorithm in the classical worm update, the worm shifting probability at a site would depend on other shifting processes at the nearest neighbor sites. The shifting probabilities at the nearest neighbor sites would further depend on the processes at the next nearest neighbor sites. Thus, it is non-trivial to write down the global balance condition in a closed form. The Metropolis scheme reduces the condition to a local form, but no room for optimization is left, except for increasing the number of possible states. In contrast, the balance condition of the worm scattering process in our approach is expressed in a closed form without using the Metropolis algorithm, as shown in the next subsection. This simple structure of the balance condition leaves much room for optimization.

\subsection{Geometric allocation approach}
\label{go}
We here detail the optimization of the transition probability in the present worm algorithm. In the MCMC method, it is crucial to optimize the transition probability for practical and efficient sampling. The problem we tackle here is how to prepare a set of appropriate transition probabilities between given states. The geometric allocation\,\cite{SuwaT2010,Suwa2014} is a versatile approach to optimizing the transition probability. The fundamental concept of this approach is that {\em the flows} between the states are purposefully allocated using a geometric graph. This geometric manner is very distinct from the conventional approaches, such as the Metropolis and the heat bath algorithms. They provide algebraic solutions that satisfy the detailed balance condition, which is a sufficient condition for the global balance. In contrast, the allocation approach converts the optimization problem into a geometric puzzle and provides a graphic solution.  Although the geometric allocation was originally introduced to break the detailed balance in Ref.\,\cite{SuwaT2010}, one of the main advantages of this approach is that we can easily arrange the transition probability in a flexible manner. We stress that the geometric allocation is not merely a representation of solutions but a versatile and efficient way to find optimal solutions.

Let us describe the rule of the puzzle game. Let $v_{ij}:=\pi_i \, p_{i\to j}$ be the raw flow from state $i$ to $j$, where $\pi_i$ is the weight, or the measure, of state $i$ apart from the normalization factor of a target distribution, and $p_{i \to j}$ is the transition probability from $i$ to $j$. Given possible states and their weights $\{\pi_i\}$, we allocate $v_{ij}$ under the two conditions: the law of probability conservation and the global balance condition, which are expressed by
\begin{equation}
  \pi_i = \sum_{j=1}^n v_{ij} \qquad \forall i
  \label{pc}
\end{equation}
and
\begin{equation}
  \pi_j = \sum_{i=1}^n v_{ij} \qquad \forall j ,
  \label{bc}
  \end{equation}
respectively, where $n$ is the number of possible states. In the worm scattering process for the square lattice Ising model, there are four possible states because a square lattice has a coordination number of four.  For example, in the case of Fig~\ref{fig:ws}, the possible states are $b$, $c$, $d$, and $e$, with $n=4$.

Let us reinterpret the conventional algorithms in this picture. It is easy to understand that the flows allocated by the Metropolis and the heat bath algorithms are represented by
\begin{equation}
  v_{ij} = \frac{1}{n-1}\min \left( \pi_i, \pi_j \right) \quad i \neq j
\end{equation}
and
\begin{equation}
  v_{ij} = \frac{\pi_i \pi_j}{ \sum_{k=1}^n \pi_k} \quad \forall i,j,  
\end{equation}
respectively. Both algorithms satisfy the detailed balance condition, which is expressed by the symmetry of the flow: $v_{ij}=v_{ji}$.

Let us set a cost function in this optimization problem. The cost function we first consider is the average rejection (worm backscattering) probability, which is given by $\sum_i v_{ii} / \sum_j \pi_j$. We optimize the flow for the average rejection probability to be minimized. This choice should be desirable because, in general, the rejection reduces the sampling efficiency of the MCMC method. Rejection minimization has also been discussed in the previous applications\,\cite{SuwaT2010,Suwa2014}.

To increase further the diffusivity of the worm head, we maximize the forward scattering probability under the condition of backscattering minimization. Our choice of the local transition probability is expected to reduce the variance of the worm length, namely the variance of the first return time for the head to come back to the tail position.

\begin{figure}
\begin{center}  
\includegraphics[width=\columnwidth,clip]{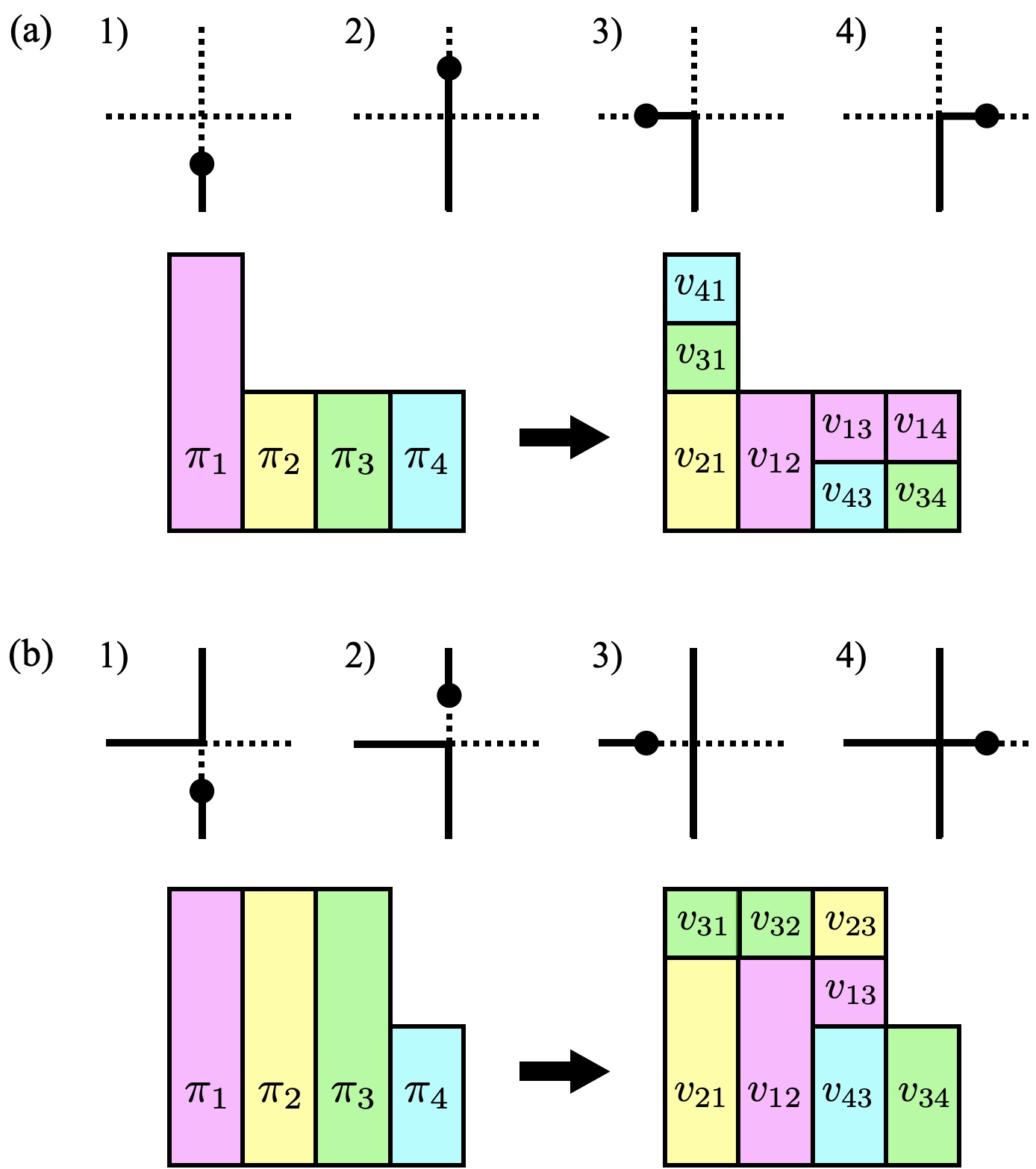}
\caption{(Color online) Geometric allocation for the square lattice Ising model. There are two cases: (a) and (b). The solid (broken) lines show the activated (deactivated) bonds, and the solid circles show the worm head. The weight, or the measure, of each state is denoted by $\pi_i$ ($i=1,2,3,4$) apart from the normalization factor of the target distribution, and the allocated raw flow from $i$ to $j$ is denoted by $v_{ij}$. The detailed balance condition is satisfied in both cases: $v_{ij}=v_{ji}$. In the case of (a), we set $v_{12}=\pi_4$, $v_{13}=v_{14}=\frac{1}{2}(\pi_1 - \pi_4)$, and $v_{34}=\frac{1}{2}(3\pi_4 - \pi_1)$. This rejection-free allocation can be performed if $3\pi_4 > \pi_1 \Leftrightarrow T < 2 / \ln 2$. In the case of (b), we set $v_{12}=\frac{1}{2}(\pi_1 + \pi_4)$, $v_{13}=v_{23}=\frac{1}{2}(\pi_1 - \pi_4)$, and $v_{34}=\pi_4$, which is possible at any temperature. The scale of the area $\pi_i$ is arbitrary; only the ratio $\pi_4 / \pi_1$ matters.}
\label{fig:go1}
\end{center}
\end{figure}

We found the optimal solution through the geometric allocation shown in Fig.~\ref{fig:go1}. Any local configuration in the worm scattering can be mapped into the case of Fig. 3 (a) or (b) through a possible rotation and flip. It is easy to confirm that Eqs.~(\ref{pc}) and~(\ref{bc}) are both satisfied: the area of each weight (color) is conserved, which is nothing but the probability conservation; the entire {\em box} shape is intact after the allocation, which guarantees the global balance. We also obtained the analytical form of the flow $v_{ij}$ corresponding to the optimal allocation, which is shown in the caption.

The rejection-free condition\,\cite{SuwaT2010} is, in general, given by
\begin{equation}
  \pi_1 \leq \sum_{i=2}^n \pi_i.
  \label{rf}
\end{equation}
This condition is equivalent to $\tanh K \geq 1/3 \iff T \leq 2 / \ln 2$ in the case of Fig.~\ref{fig:go1}\,(a) and always satisfied in the case of Fig.~\ref{fig:go1}\,(b). Here, the ratio $\pi_4 / \pi_1(=\tanh K)$ depends on the temperature in the simulation. Our update is rejection free for $T \leq 2 / \ln 2 \simeq 2.885$, including the critical temperature $T_{\rm c}=2/\ln(1+\sqrt{2}) \simeq 2.269$\,\cite{KramersW1941}. In addition, the forward scattering probability $(v_{12}+v_{21}+v_{34}+v_{43})/\sum_j \pi_{j}$ is maximized under rejection minimization in both cases.

We chose the unique solution satisfying the detailed balance condition under backscattering minimization and forward scattering maximization. Technically, directed worm scattering always breaks the detailed balance in the extended state space. Nevertheless, if local worm scattering satisfies the detailed balance condition without taking the direction into account, the whole worm update from insertion to removal ensures the detailed balance in the original state space\,\cite{SyljuasenS2002}. It is easy to find many (actually infinite) solutions to satisfy the required conditions [Eqs.~(\ref{pc}) and~(\ref{bc})] thanks to the geometric picture. Even solutions breaking detailed balance can be readily found\,\cite{SuwaT2010}. For example, starting from the solution shown in Fig.~\ref{fig:go1}\,(a), we can increase a certain amount of $v_{13}$, $v_{34}$, and $v_{41}$, while decreasing the same amount of $v_{31}$, $v_{43}$, and $v_{14}$. This modified solution again satisfies Eqs.~(\ref{pc}) and~(\ref{bc}) without the detailed balance because $v_{ij} \neq v_{ji}$ for $(i,j)=(1,3),(1,4),(3,4)$. The modified solution, as well as the original solution [Fig.~\ref{fig:go1}\,(a)], has the minimized (zero) backscattering rate $(= v_{11} + v_{22} + v_{33} + v_{44})/\sum_j \pi_{j}$ and the maximized forward scattering rate $(= v_{12} + v_{21} + v_{34} + v_{43})/\sum_j \pi_{j}$. If local flows break the detailed balance condition ($v_{ij} \neq v_{ji}$), the whole worm update breaks the detailed balance in the original state space as well, which is called irreversible. Although it is possible to improve the efficiency by breaking detailed balance, we have not yet found any irreversible solution that works significantly better than the present choice in the case of the Ising model. We selected the present reversible solution because it is unique and easy to prove the ergodicity (discussed below in Sec.~\ref{er}). Irreversible Markov chains, nevertheless, have the potential to play an essential role in Monte Carlo dynamics.

\begin{figure*}
\begin{center}  
\includegraphics[width=2\columnwidth]{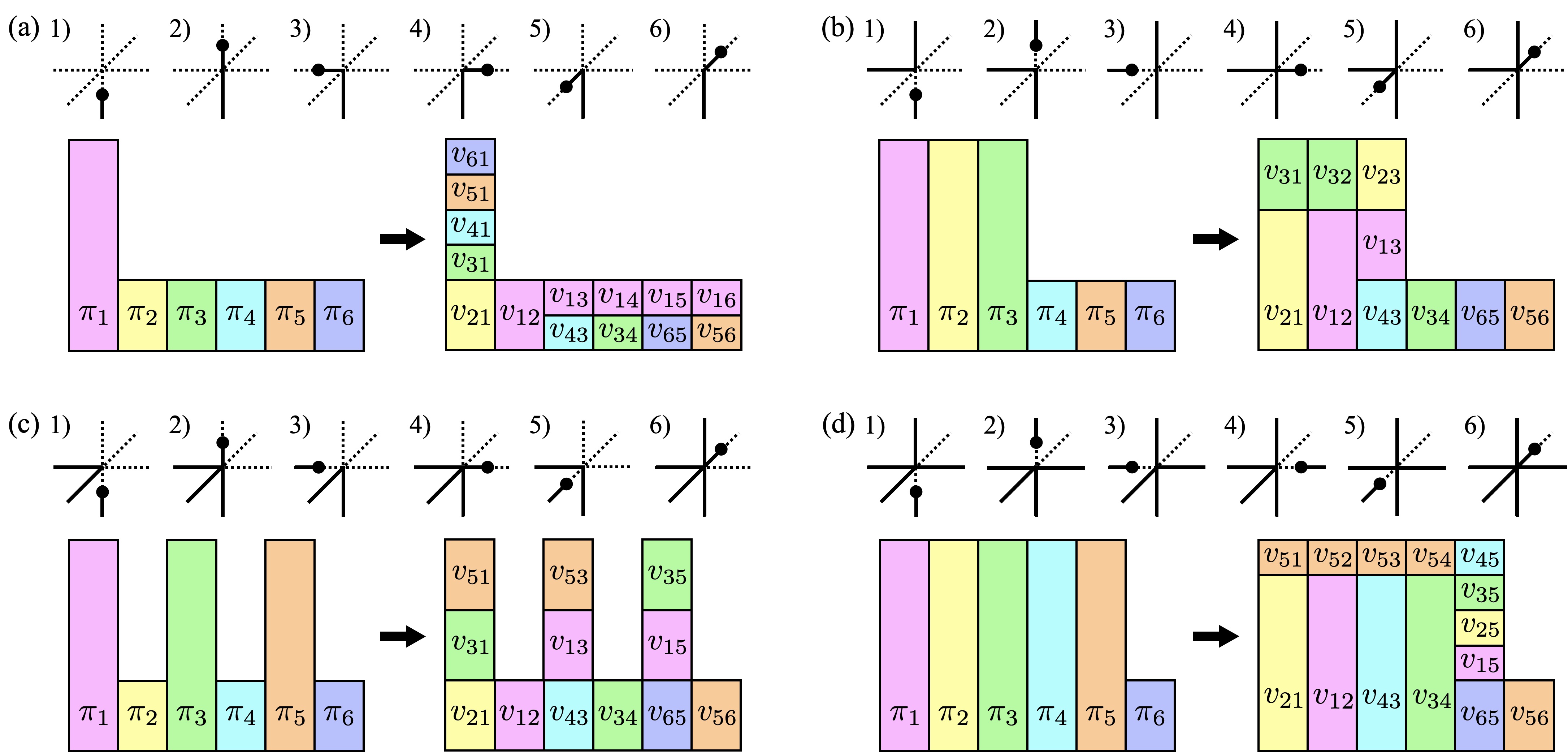}
\caption{(Color online) Geometric allocation for the simple cubic lattice Ising model. There are four cases: (a), (b), (c), and (d). The six possible states are indexed such that (1,\,2), (3,\,4), and (5,\,6) are the pairs of the states from and to which the worm head forward scatters like the square lattice case. The detailed balance condition is satisfied in all the cases: $v_{ij}=v_{ji}$. In the case of (a), we set $v_{12}=\pi_6$, $v_{13}=v_{14}=v_{15}=v_{16}=\frac{1}{4}(\pi_1 - \pi_6)$, and $v_{34}=v_{56}=\frac{1}{4}(5\pi_6 - \pi_1)$; in (b), $v_{12}=\frac{1}{2}(\pi_1 + \pi_6)$, $v_{13}=v_{23}=\frac{1}{2}(\pi_1 - \pi_6)$, and $v_{34}=v_{56}=\pi_6$; in (c), $v_{12}=v_{34}=v_{56}=\pi_6$ and $v_{13}=v_{15}=v_{35}=\frac{1}{2}(\pi_1 - \pi_6)$; in (d), $v_{12}=v_{34}=\frac{1}{4}(3\pi_1 + \pi_6)$, $v_{15}=v_{25}=v_{35}=v_{45}=\frac{1}{4}(\pi_1 - \pi_6)$, and $v_{56}=\pi_6$.  This rejection-free allocation in (a) can be performed if $5\pi_6 > \pi_1 \Leftrightarrow T < 2 / \ln (3/2)$. The allocations in (b), (c), and (d) are possible at any temperature.}
\label{fig:go3}
\end{center}
\end{figure*}

We can calculate all the transition probabilities, $p_{i \to j}= v_{ij} / \pi_i \,\forall i,j$, before simulation and prepare a look-up table storing the probabilities. In the actual simulations, we choose the next state in each worm scattering by using Walker's method of alias\,\cite{FukuiT2009,HoritaST2017}. The advantage of Walker's method is that the computation time, which is $O(1)$, does not increase with the number of possible states $n$ in contrast to the computation time of a simple binary search, which is $O(\log n)$. The present algorithm needs no extra computational cost, compared to the classical worm algorithm.

In the simple cubic lattice case, we chose a set of flows, as illustrated in Fig.~\ref{fig:go3}. The six possible states are indexed such that (1,\,2), (3,\,4), and (5,\,6) are the pairs of the states from and to which the worm head forward scatters like the square lattice case. The allocation patterns depending on the local configuration are all shown in Fig.~\ref{fig:go3}. We express the analytical form of $v_{ij}$ as well in the caption. The rejection-free condition~[Eq.~(\ref{rf})] is satisfied for $T \leq 2/\ln(3/2) \simeq 4.933$, including the critical temperature $T_{\rm c} \approx 4.511$\,\cite{DengB2003}. In addition to the conditions of backscattering minimization and forward scattering maximization, we here impose an additional condition to find the unique solution; the variance of the forward scattering flow, $\sum_{k=1,3,5} ( v_{k \, k+1} - \overline{v})^2$, where $\overline{v}= \frac{1}{3} \sum_{k=1,3,5} v_{k \, k+1}$, is minimized. Other solutions, nevertheless, are expected to work as well as our choice does as long as the backscattering probability is minimized, and the forward scattering probability is maximized.

Our geometric allocation approach to optimizing the worm scattering probability can be generalized to many physical models, such as the $|\phi|^4$ model, the Potts model, the O($n$) loop model, and lattice QCD. It is expected to improve the computational efficiency of the worm update for these models as well as for the Ising model.

\subsection{Ergodicity}
\label{er}
We here show that the Markov chain created by the present worm algorithm is (uniformly) ergodic in the extended state space; equivalently, it is irreducible and aperiodic\,\cite{RobertC2004}. Any configuration under the loop constraint is represented by a combination of loops formed by activated bonds. The worm can create any loop with a finite probability. (Note that the forward scattering probability is always positive in the present flow allocation.) Hence any state with and without kinks can be visited from the vacuum state, in which $n_b=0$ $\forall b$. Because our solution of the transition probability holds the detailed balance, any two states in the extended state space are connected by the transition kernel; the Markov chain is irreducible.

Let us next consider the aperiodicity. Even if never backscatters, the worm can come back to the same physical state (with no kink). There are many paths for the kink to start from and end at the vacuum state. For example, the paths formed by nine and 11 worm scattering steps (going around a plaquette twice) exist for both the square and cubic lattice models. Let $p$ and $q$ be nine and 11, respectively. Because $p$ and $q$ are coprime, B\'{e}zout's identity states $\exists a,b \in \mathbb{Z} \mbox{ s.t. } ap+bq=\mbox{gcd}(p,q)=1$. We can choose $a$ and $b$ such that $-q<a<0$ and $0<b<p$.

We prove $\forall n \geq pq-1$, $\exists c,d \in \mathbb{N} \mbox{ s.t. } n=cp+dq$. First, we can express $pq-1=-ap+(p-b)q$, where $-a>0$ and $p-b>0$. We here use mathematical induction: if $\exists m \in \mathbb{N} \mbox{ s.t. } m \geq pq-1$ and $\exists \alpha,\beta \in \mathbb{N} \mbox{ s.t. } m=\alpha p + \beta q$, then $\exists \alpha', \beta' \in \mathbb{N} \mbox{ s.t. } m+1=\alpha' p + \beta' q$. We can express $m+1=(\alpha+a)p+(\beta+b)q=(\alpha+a+q)p+(\beta+b-p)q$. If $\alpha+a \geq 0$, simply $\alpha'=\alpha+q$ and $\beta'=\beta+b$. If $\alpha+a < 0$, we can take $\alpha'=\alpha+a+q \geq 0$ and $\beta'=\beta+b-p \geq 0$. The last inequality follows from $\alpha+a<0 \Rightarrow \alpha p + ap < 0 \Rightarrow m+1-(\beta+b)q<0 \Rightarrow m+1 - (\beta+b-p)q < pq \Rightarrow (\beta+b-p)q > m + 1 - pq \geq 0 \Rightarrow \beta+b-p \geq 0$.

Therefore, the vacuum state can be revisited from itself with a finite probability after $n \geq pq-1$ worm scattering steps: the vacuum state is aperiodic in the extended space. Hence, the Markov chain is aperiodic.

We can choose an irreversible solution instead of the present reversible one, as mentioned in Sec.~\ref{go}. It is not trivial to prove the ergodicity of irreversible Markov chains. Nevertheless, we have tested a couple of irreversible solutions and confirmed that the results are consistent. We thus expect many solutions to generate ergodic Markov chains even without detailed balance.

\begin{figure}
\begin{center}  
\includegraphics[width=\columnwidth]{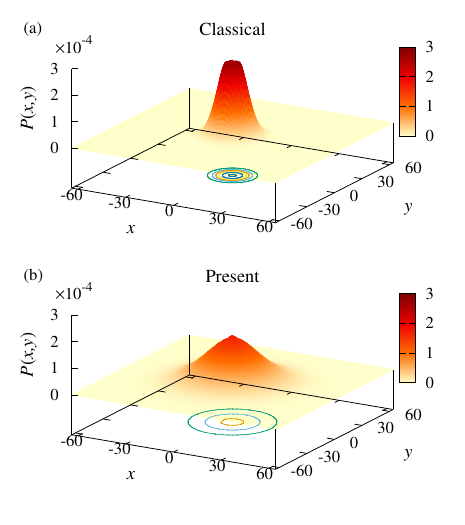}
\caption{(Color online) Probability distributions of the difference between the coordinates of the two kinks (head and tail) for the $L=128$ square lattice model at the critical temperature in (a) the classical and (b) the present worm updates. The coordinates were measured 256 worm shifting or scattering steps after the insertion. The contours show the coordinates at which $P(x,y)=0.5, 1.0, 1.5, 2.0$, and $2.5 \times 10^{-4}$. Because the two kinks are removed when meeting each other, the distribution is lowered near the center, which is more significant in the classical worm update. The removed worms are not shown here but counted in the normalization.}
\label{fig:P}
\end{center}
\end{figure}
\begin{figure}
\begin{center}  
\includegraphics[width=\columnwidth]{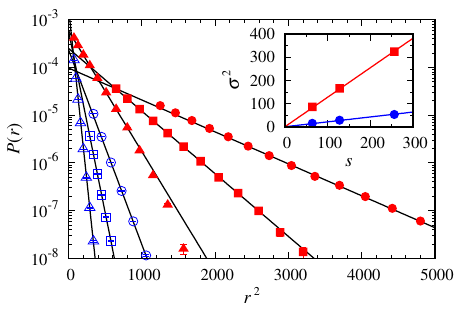}
\caption{(Color online) Tails of the probability distributions of the distance between the two kinks in the classical (open) and the present (solid) worm updates 64 (triangles), 128 (squares), and 256 (circles) local worm steps after the insertion, measured in the $L=128$ square lattice Ising model at the critical temperature. The distribution was measured at $r=|{\mathbf r}|$, where ${\mathbf r = (x,y)}$ and $|x|=|y|$. The tails are fitted to Gaussian distributions: $ P(r) \propto e^{-r^2 / 2 \sigma^2 }$, where $\sigma^2$ is a parameter (variance). The inset shows the linear scaling of the estimated variance in the classical (circles) and the present (squares) worm updates as a function of the number of worm steps ($s$). The variance in the present worm update is approximately six times as large as in the classical worm update.}
\label{fig:P_r}
\end{center}
\end{figure}

\subsection{Enhanced diffusivity}
\label{dist}
We demonstrate here that the present worm algorithm indeed enhances the diffusivity of the worm head. Figure~\ref{fig:P} shows the probability distribution of the difference between the coordinates of the two kinks (the worm head and tail) in the classical and the present worm updates for the $L=128$ square lattice Ising model at the critical temperature. The coordinates were measured 256 worm shifting or scattering steps after the insertion. The distribution in the present worm update is much broader than in the classical worm update. The removed worms at $(x,y)=(0,0)$ before 256 worm steps are not shown in Fig.~\ref{fig:P} but counted in the normalization.

The distribution tail of the kink distance is well approximated by a Gaussian distribution, as shown in Fig.~\ref{fig:P_r}. We estimated the variances of the Gaussian distributions 64, 128, and 256 local worm steps after the insertion. In Fig.~\ref{fig:P_r}, although some faster decay is observed in the distribution after 64 scattering steps of the present worm update, the tails of the distributions after 128 and 256 steps are well fitted to Gaussian distributions up to longer distances. We found a linear growth of the variance as a function of the number of local worm steps, as shown in the inset. The variance in the present worm update is six times as large as in the classical worm update. These observations indicate that the present method successfully enhances the diffusivity of the worm head, which is expected to improve sampling efficiency.

\subsection{Estimators}
\label{obs}
Many physical quantities can be measured in the present worm simulation as well as in the classical worm simulation. For example, the total energy can be estimated by the same observable~[Eq.~(\ref{e})]. Nevertheless, estimators associated with the extended state space need to be slightly modified. As relevant quantities, we here explain how to measure the spin correlation function and the magnetic susceptibility.

In the classical worm algorithm, the state space is extended to include the configurations that contain (up to) two kinks at sites of a lattice. How many times the two kinks are at sites $i$ and $j$ directly contributes to the estimator of the spin correlation between sites $i$ and $j$, as mentioned in Sec.~\ref{alg1}. On the other hand, the present worm is never located at the sites during the update processes.

Let us here consider a {\it virtual} process of shifting the two kinks from bonds to adjacent sites of the bonds. There are four choices of sites because each bond connects two sites. We choose a pair of sites at random, one from the two adjacent sites of the head and the other from the two adjacent sites of the tail. We then consider using the Metropolis algorithm to accept or reject the virtual shift. The acceptance probability depends on the change of $\ell = \sum_b n_b$. If this virtual shift were accepted, we would count one for measuring the associated spin correlation in a manner similar to the classical worm algorithm. We next consider a reverse process of shifting the kinks from the sites to the original bonds, using the Metropolis algorithm again. If this reverse shift were rejected, we would count one for measuring the spin correlation again. We would repeat the reverse shifting process and continue counting one while the kinks would be on sites. The average count through these virtual processes is given by the ratio of the weight of a site-kink configuration to the weight of a bond-kink configuration. Thus, we can use the weight ratio as the reweighting factor from a bond-kink configuration to a site-kink configuration. Since we assume each kink to be at the center of a bond in the present algorithm, the reweighting factor is given by a simple form. To calculate the magnetic susceptibility, we take the average over the four possible choices and sum up the reweighting factors during the worm scattering process from insertion to removal without going through the virtual processes.

From the above argument, a magnetic susceptibility estimator in the present algorithm is given by
\begin{equation}
  \hat{\chi} = \frac{\beta}{4z w} \sum_{\rm path} f_{\rm rew}, \label{chi}
\end{equation}
where $\beta$ is the inverse temperature, $z$ is the coordination number (four for a square lattice and six for a cubic lattice), $w$ is the extra weight the worm carries,
\begin{equation}
  f_{\rm rew} = \left( s + \frac{1}{s} \right) f_h \label{f-rew}
\end{equation}
is the reweighting factor after a worm scattering process, and $s \equiv \sqrt{\tanh K}$. In Eq.~(\ref{f-rew}), $f_h$ is $2/s$ if the head is on an activated bond, $2s$ on a deactivated bond, and $s+\frac{1}{s}$ on a half-activated and half-deactivated bond. In other words, $f_h$ takes $2/s$ or $2s$ if the head comes back to the tail. It takes $s+\frac{1}{s}$ otherwise. The summation in Eq.~(\ref{chi}) means that $f_{\rm rew}$ is calculated after each worm scattering process and summed over the scattering processes in step 2. The sum of the reweighting factors is divided by four because we take the average of the four reweighting factors depending on the choice of adjacent sites. Furthermore, it is divided by the coordination number because of the multiple counts in the virtual shift from bonds to sites.

Let us consider the extra weight~$w$ in Eqs.~(\ref{chi}) and~(\ref{chi-sim-ell}). In the worm algorithm, we can arbitrarily set the weight of each state in the extended space: we assume that the present worm has extra weight, a factor of a half, to insert and remove the worm with probability one in the Ising model. The value of the extra weight comes from the fact that there are two possible directions for the worm head to go in. We need to take this extra weight into account for estimators related to the extended space. In the susceptibility estimator~(\ref{chi}), the reweighting factor needs to be divided by the extra weight ($w=1/2$ for the Ising model). It is straightforward to calculate other quantities, such as the Fourier transformed correlation function and the correlation length. Note that while being a factor of a half for the Ising model, the extra weight that the worm carries may depend on models and worm variants used in the simulation.

We easily find
\begin{equation}
  \chi \sim \frac{ \beta  }{ 4 z w } \left( s + \frac{1}{s} \right)^2 \langle \ell_{\rm worm} \rangle,
  \label{chi-sim-ell}
\end{equation}
where $\ell_{\rm worm}$ is the worm length. It is because $f_h$ takes
$s+\frac{1}{s}$ unless the head and tail are located at the same position. This estimation is comparable to Eq.~(\ref{chi-cl}) in the case of the classical worm algorithm.

\subsection{Avoiding bias}
\label{cwa}
Before closing this section on the methodology, we discuss a possible bias introduced by the fixed time simulation of the worm algorithm and the Wolff cluster algorithm\,\cite{Wolff1989}. In these methods, the computation time for a Monte Carlo step depends on the worm length or the cluster size (in the Wolff algorithm). The mean worm length, which is approximately proportional to the magnetic structure factor ($=\chi / \beta$) as shown in Eqs.~(\ref{chi-cl}) and~(\ref{chi-sim-ell}), is usually a decreasing function of temperature and indeed so in the present Ising models. In contrast, the total energy is an increasing function of temperature. When the configuration in the simulation is a higher energy configuration, the computation time for the subsequent Monte Carlo step (from worm insertion to removal) will be shorter on average. In other words, the needed computational time to sample a high energy configuration is shorter on average than to sample a low energy configuration. As a result, given a simulation time, say, one hour, high energy configurations tend to be sampled more often than low energy configurations. Therefore, such a fixed time simulation creates a bias. For example, if a parallel simulation is run for a certain period using independent Markov chains, an estimator that naively averages over chains has a bias. To avoid this bias, we need to fix the total Monte Carlo steps for each chain instead of the total run time and take the average over chains that run the same Monte Carlo steps. The bias we discuss here can be caused in the worm algorithm for quantum systems as well. Although the bias might be tiny, we carefully run simulations avoiding it.

\section{How to compare MCMC samplers}
\label{mc}
We discuss how to quantify the computational efficiency of the MCMC sampler. There are mainly two points to consider\,\cite{Suwa2014}: the relaxation rate and the sampling efficiency. In the former, as Monte Carlo samples are taken after the thermalization, faster relaxation to a target distribution allows for sampling from an earlier Monte Carlo step; in the latter, more efficient sampling yields a smaller statistical error. The mean squared error of an estimator is proportional to the inverse of the number of samples (Monte Carlo steps) according to the central limit theorem\,\cite{RobertC2004}. The sampling efficiency of the MCMC update should be quantified by the prefactor of the scaling, that is, the asymptotic variance\,\cite{Suwa2014}. We explain here how to measure relevant quantities in the present method.

The relaxation rate is quantified by the exponential autocorrelation time. The autocorrelation function exponentially decays in large Monte Carlo steps, which is the case for the finite size systems we study in the present paper. We calculate the function by running independent simulations and estimate the exponential autocorrelation time as a fitting parameter. In the present paper, we use a single exponential as the fitting function and estimate the error bar of the fitting parameter by using bootstrapping\,\cite{DavisonH1997,SenSS2015}.

In the worm algorithm, we consider each Monte Carlo step to be a one-time worm update from insertion to removal. In other words, the number of Monte Carlo steps is equal to how many times the head comes back to the tail. Here, the number of Monte Carlo steps should be measured in units of the number of sites for a fair comparison. An autocorrelation time $\tau'_{\rm exp}$ estimated by fitting to an exponential function is rescaled:
\begin{align}
  \tau_{\rm exp} = \tau'_{\rm exp} \frac{ \la \ell_{\rm worm} \ra }{N} ,
  \label{tau_exp_worm}
\end{align}
where $ \la \ell_{\rm worm} \ra$ is the mean worm length, and $N$ is the number of sites. The mean worm length differs for the classical and the present worm updates as the state space is extended in different manners.

The sampling efficiency of the MCMC method is related to the integrated
autocorrelation time. It can be estimated by the relation
\begin{equation}
  \tau_{\rm int}' = \frac{\sigma^2}{2 \bar{\sigma}^2} ,
  \label{tau_int_est}
\end{equation}
where $\sigma^2$ is the mean squared error, namely the square of the statistical error, calculated by binning analysis using a much larger bin size than the exponential autocorrelation time, and $\bar{\sigma}^2$ is calculated without binning. The above estimator~(\ref{tau_int_est}) gives the exact integrated autocorrelation time~(\ref{tau_int}) in the limit of large number of Monte Carlo steps\,\cite{LandauB2005}. In a manner similar to Eq.~(\ref{tau_exp_worm}), it is rescaled:
\begin{align}
  \tau_{\rm int} = \tau'_{\rm int} \frac{ \la \ell_{\rm worm} \ra }{N}.
  \label{tau_int_worm}
\end{align}

Although the integrated autocorrelation time is useful for studying Monte
Carlo dynamics, we stress that the sampling efficiency of the Monte Carlo method should be quantified by the asymptotic variance, which is the prefactor of the asymptotic scaling:
\begin{equation}
  \sigma_{\hat{\mathcal O}}^2 \approx \frac{ v_{{\rm asymp},{\hat{\mathcal O}}} }{M} \label{clt},
\end{equation}
where $\sigma_{\hat{\mathcal O}}^2$ is the mean squared error of an estimator $\hat{\mathcal O}$, $v_{{\rm asymp}, {\hat{\mathcal O}}}$ is the asymptotic variance of $\hat{\mathcal O}$, and $M$ is the renormalized number of Monte Carlo steps. Here we assume $\hat{\mathcal O}$ to be an unbiased estimator of a physical quantity ${\mathcal O}$: $\la \hat{\mathcal O} \ra = {\mathcal O}$. Then the asymptotic variance is represented by
\begin{equation}
  v_{{\rm asymp},{\hat{\mathcal O}}} = 2 \tau_{{\rm int}, {\hat{\mathcal O}}} v_{\hat{{\mathcal O}}} \label{v_asymp},
  \end{equation}
where $v_{\hat{\mathcal O}}= \la \hat{\mathcal O}^2 \ra - \la \hat{\mathcal O} \ra ^2$ is the variance of $\hat{\mathcal O}$.

In the present paper, according to Eqs.~(\ref{tau_int_est}), (\ref{tau_int_worm}), (\ref{clt}) and~(\ref{v_asymp}), we estimate the variances using the jackknife method\,\cite{Berg2004} and the following relations:
\begin{align}
      v_{{\rm asymp},\hat{\mathcal O}} &= M' \frac{\sigma_{\hat{\mathcal O}}^2}{\mu_{\hat{\mathcal O}}^2} \frac{ \la \ell_{\rm worm} \ra }{N} \label{v_asymp_est}\\
      v_{\hat{\mathcal O}} &= M' \frac{\bar{\sigma}_{\hat{\mathcal O}}^2}{\mu_{\hat{\mathcal O}}^2}, \label{v_obs_est}
\end{align}
where $M'$ is the original number of Monte Carlo steps used for sampling in a simulation, $\sigma_{\hat{\mathcal O}}^2$ and $\bar{\sigma}_{\hat{\mathcal O}}^2$ are the mean squared errors of an estimator $\hat{\mathcal O}$ with and without binning, and $\mu_{\hat{\mathcal O}}$ is the average of the samples, respectively. We here use the squared coefficient of variation ($\sigma^2/\mu^2$) to remove a trivial dependence on the definition of the estimator: for example, the variances of the total energy and the energy density are identical.

The renormalization of the number of Monte Carlo steps is necessary also for the Wolff algorithm. We simply replace the mean worm length with the mean cluster size in Eqs.~(\ref{tau_exp_worm}), (\ref{tau_int_worm}), and (\ref{v_asymp_est}).

\begin{figure}
\begin{center}  
\includegraphics[width=\columnwidth]{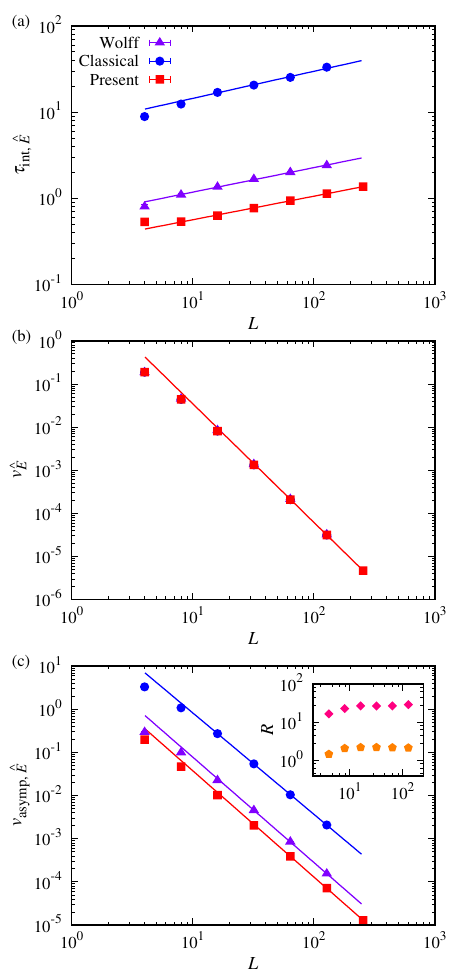}
\caption{(Color online) (a) The integrated autocorrelation time, (b) the variance, and (c) the asymptotic variance of the total energy estimator as a function of the system length of the simple cubic lattice Ising model at the critical temperature. The exponents of $\tau_{{\rm int}, \hat{E}}$ are estimated to be $0.28$, $0.31$, and $0.27$, and those of $v_{{\rm asymp}, \hat{E}}$ are to be $-2.44$, $-2.35$, and $-2.46$ in the Wolff cluster (triangles), the classical worm (circles), and the present worm (squares) updates, respectively; the exponent of $v_{\hat{E}}$ is estimated to be $-2.75$ in all the updates. The inset of panel (c) shows the ratios of the asymptotic variance in the classical worm (diamonds) and the Wolff cluster (pentagons) updates to the one in the present worm update. They are approximately 27 and 2.2 for large system sizes, respectively.}
\label{fig:e-3D}
\end{center}
\end{figure}

\section{Results}
\label{result}
We investigate the performance of our worm algorithm for the simple cubic lattice Ising model, focusing on critical slowing down at the transition temperature.  We compare the present algorithm with the classical worm\,\cite{ProkofievS2001} and the Wolff algorithms\,\cite{Wolff1989}. The ensemble used in the simulations is represented by Eq.~({\ref{Z}) at the critical temperature, $1/T_{\rm c} \approx 0.22165455$\,\cite{DengB2003}. Periodic boundaries were used in all the spatial directions. We optimize the worm scattering probability, as illustrated in Fig.~\ref{fig:go3}. More than $2^{24}$ Monte Carlo samples were taken, in total, after $2^{16}$ thermalization steps.

For a fair comparison, we adopt $N$ local worm processes in the worm algorithms and $N$ spin updates in the Wolff algorithm as the unit of time in the Monte Carlo dynamics. Here $N$ is the number of sites of the system. The autocorrelation times were rescaled as shown in Eqs.~(\ref{tau_exp_worm}) and~(\ref{tau_int_worm}). The mean worm length in the classical worm update is proportional to the magnetic susceptibility: $\la \ell_{\rm classical \ worm} \ra = \chi / \beta \propto L^{\gamma/\nu}$, where $\gamma$ and $\nu$ are the critical exponents of the susceptibility and the correlation length, respectively\,\cite{LandauB2005}. We found a relation between the worm lengths in the present and the classical worm updates: $\la \ell_{\rm present \ worm} \ra \approx 1.765 \la \ell_{\rm classical \ worm} \ra$ for $L \geq 16$.

The integrated autocorrelation time, the variance, and the asymptotic variance of the energy estimator are shown in Fig.~\ref{fig:e-3D}. We calculated these quantities in the manner explained in Sec.~\ref{mc}. Using the Wolff algorithm, we calculated the total energy from the spin configuration. The present algorithm produces the shortest integrated autocorrelation time and the smallest asymptotic variance. The shorter correlation time in the present worm update allowed us to run simulations for the larger system size.

Fitting a power law to data, we estimate the exponents of $\tau_{{\rm int}, \hat{E}}$ to be $0.28$, $0.31$, and $0.27$, and those of $v_{{\rm asymp}, \hat{E}}$ to be $-2.44$, $-2.35$, and $-2.46$ in the Wolff cluster (triangles), the classical worm (circles), and the present worm (squares) updates, respectively; we estimate the exponent of $v_{\hat{E}}$ to be $-2.75$ in all the updates. We expect the three algorithms to produce the same exponent asymptotically. Nevertheless, as shown in the inset of Fig.~\ref{fig:e-3D}\,(c), the asymptotic variance in the present worm update is approximately 27 and 2.2 times as small as in the classical worm and the Wolff cluster updates, respectively.

In the worm algorithm, the weight of the loop configuration is not the Boltzmann distribution, as shown in Eq.~(\ref{Z}): $\pi_i \propto (\tanh K)^{\ell_i}$. Thus, estimators for a physical quantity naturally depend on the representation. We can construct an estimator in the worm algorithm whose mean value is identical to a physical quantity of the original spin system. Nevertheless, the variances of the estimators are generally different. The variance of the energy estimator is the same for the classical and the present worm algorithms simply because the same estimator is used. However, the variance is different from the one in the Wolff algorithm: the estimators are different, although their mean values are identical. Nevertheless, the difference is small, and both the variances show almost the same exponent ($\approx -2.75$) of the power-law decay, as shown in Fig.~\ref{fig:e-3D}\,(b). The variance of the energy estimator in the Wolff algorithm is nothing but the energy variance of the original spin system, which is proportional to the specific heat. The asymptotic scaling, therefore, should be $v_{\hat{E}} \propto L^{\alpha/\nu - d}$ with the exponent $\alpha/\nu - d \approx -2.826$\,\cite{DengB2003}. The present estimate is slightly larger by 2.7\%. Data of larger system sizes seem to be needed for a single power-law fit we use to match the exponent estimated from the more sophisticated analysis.

\begin{figure}
\begin{center}  
\includegraphics[width=\columnwidth]{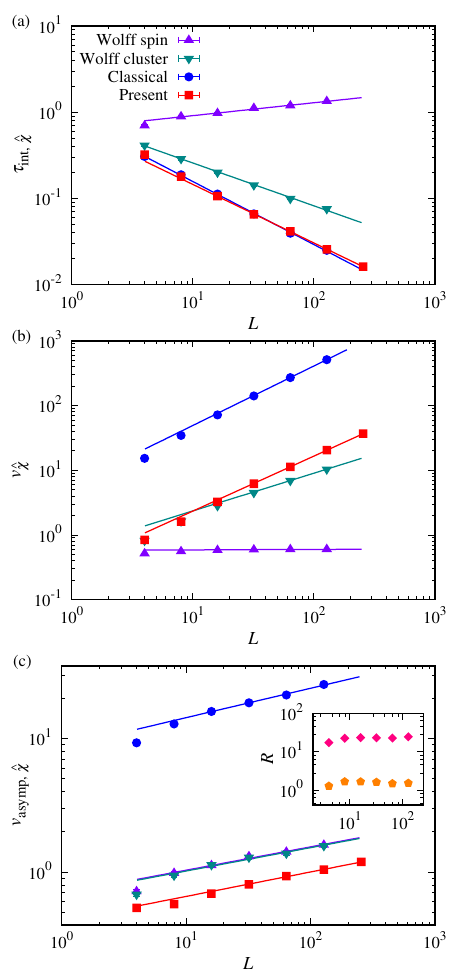}
\caption{(Color online) (a) The integrated autocorrelation time, (b) the variance, and (c) the asymptotic variance of the magnetic susceptibility estimator as a function of the system length of the simple cubic lattice Ising model at the critical temperature in the Wolff (triangles), the classical (circles), and the present worm (squares) algorithms. In the Wolff algorithm, we test two estimators using the spins (Wolff spin) and the cluster size (Wolff cluster) (see the main text for the detail of the estimators). The exponents of $\tau_{{\rm int}, \hat{\chi}}$ are estimated to be $0.150(9)$, $-0.50(1)$, $-0.731(7)$, and $-0.679(4)$, those of $v_{\hat{\chi}}$ are to be $0.01$, $0.58$, $0.92$, and $0.85$, and those of $v_{{\rm asymp}, \hat{\chi}}$ are to be $0.18$, $0.18$, $0.22$, and $0.18$ in the Wolff spin, in the Wolff cluster, in the classical worm, and in the present worm updates, respectively. The inset of panel (c) shows the ratios of the asymptotic variance in the classical worm (diamonds) and the Wolff cluster (pentagons) updates to the one in the present worm update, which are approximately 23 and 1.6 for large system sizes, respectively.}
\label{fig:sus-3D}
\end{center}
\end{figure}

The quantities of the magnetic susceptibility estimators are shown in Fig.~\ref{fig:sus-3D} like the energy estimator. In the Wolff algorithm, we test two estimators: $\hat{\chi} = \beta M_z^2 /N$ (here dubbed Wolff spin), where $M_z$ is the total magnetization of spins, and $\hat{\chi} = \beta \ell_{\rm cl} $ (Wolff cluster), where $\ell_{\rm cl} $ is the cluster size. We estimate the exponents of $\tau_{{\rm int}, \hat{\chi}}$ to be $0.150(9)$, $-0.50(1)$, $-0.731(7)$, and $-0.679(4)$, those of $v_{\hat{\chi}}$ to be $0.01$, $0.58$, $0.92$, and $0.85$, and those of $v_{{\rm asymp}, \hat{\chi}}$ to be $0.18$, $0.18$, $0.22$, and $0.18$ in the Wolff spin, in the Wolff cluster, in the classical worm, and in the present worm updates, respectively. The numbers in the parentheses indicate the statistical error, one standard deviation, in the preceding digit. Interestingly, while the exponents of $\tau_{{\rm int}, \hat{\chi}}$ and $v_{\hat{\chi}}$ are different for each estimator and algorithm, the exponent of $v_{{\rm asymp}, \hat{\chi}}$ is most likely identical. Particularly, $v_{{\rm asymp}, \hat{\chi}}$ is almost the same for the two estimators in the Wolff algorithm. Nonetheless, the asymptotic variance in the present worm update is approximately 23 and 1.6 times as small as in the classical worm and the Wolff cluster updates, respectively, as shown in the inset of Fig.~\ref{fig:sus-3D}\,(c).

We note that the susceptibility estimator is different for each case. Although the variance of the Wolff-spin estimator (simply using spins) includes four spin correlations, the variances of the estimators in the worm algorithms do not. As we mentioned above, this is because the estimators in the worm updates [Eq.~(\ref{chi-cl}) and Eq.~(\ref{chi})] are different from the Wolff-spin estimator. In practice, the variances were measured using Eq.~(\ref{v_obs_est}).

We emphasize that the sampling efficiency of the Monte Carlo method should be quantified by the asymptotic variance. As shown in Fig.~\ref{fig:sus-3D}, the exponent of the integrated autocorrelation time in the classical worm update is much smaller than in the Wolff cluster update, but the exponent of the variance in the classical worm update is much larger than in the Wolff cluster update. Interestingly, the exponent of the asymptotic variance is almost the same for the two algorithms. Indeed, the asymptotic variance in the classical worm update is much larger than in the Wolff cluster update.

The present worm update successfully reduces the variance of the susceptibility estimator. Because the worm length is proportional to the susceptibility exactly in the classical worm update as shown in Eq.~(\ref{chi-cl}) and approximately in the present worm update as shown in Eq.~(\ref{chi-sim-ell}), the variance of the worm length is also significantly reduced by the present algorithm. We expect the overall performance improvement to be attributed to the variance reduction of the worm length.

\begin{figure}  
\begin{center}
\includegraphics[width=\columnwidth]{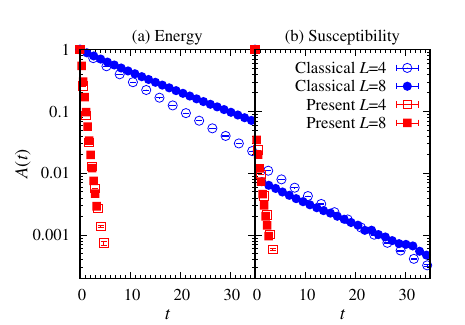}
\caption{(Color online) Autocorrelation functions of (a) the total energy and (b) the magnetic susceptibility estimators in the classical (circles) and the present (squares) worm updates for the $L=4$ (open) and $8$ (solid) simple cubic lattice Ising model. The horizontal axis is the rescaled time of the Monte Carlo dynamics in units of $L^3$ worm shifting or scattering steps.}
\label{fig:A}
\end{center}
\end{figure}

We investigate the relaxation rate as well as the sampling efficiency of the present update. The autocorrelation functions~(\ref{A}) of the total energy and the magnetic susceptibility estimators for $L=4$ and $8$ are shown in Fig.~\ref{fig:A}, calculated from more than $2^{30}$ independent Markov chains (sample paths). Each chain was sampled after thermalization steps that are much longer than the obtained exponential autocorrelation time, so the dependence on the initial state in the simulations is negligible in the present results. The function of the energy estimator shows an almost single exponential decay; that of the susceptibility estimator shows some fast and slow decays. While $\tau_{{\rm int}, \hat{\chi}}$ in the classical worm update decreases with $L$ as shown in Fig.~\ref{fig:sus-3D}\,(a), $\tau_{{\rm exp}, \hat{\chi}}$ for $L=8$ is larger than for $L=4$ as shown in Fig.~\ref{fig:A}\,(b). The reason why $\tau_{{\rm int}, \hat{\chi}}$ decreases with $L$ in contrast to $\tau_{{\rm exp}, \hat{\chi}}$ is that the prefactor of the slow mode decreases with $L$, which is also seen in Fig.~\ref{fig:A}\,(b).

We show the exponential autocorrelation times as a function of $L$ in Fig.~\ref{fig:tau_exp-3D}. The bootstrap method was used in the estimation of the fitting parameter as mentioned in Sec.~\ref{mc}. We found $\tau_{{\rm exp}, \hat{\chi}} \approx \tau_{{\rm exp}, \hat{E}}$, which is most likely the maximum exponential autocorrelation time among all the estimators. In addition, the autocorrelation function of the energy estimator is well approximated by a single exponential function, as shown in Fig.~\ref{fig:A}\,(a). Thus, the exponential and the integrated autocorrelation times should be almost the same: $\tau_{{\rm exp}, \hat{E}} \approx \tau_{{\rm int}, \hat{E}}$, which we indeed confirmed in the present results. We hence found the asymptotic scaling: $\tau_{{\rm exp}, \hat{\chi}} \approx \tau_{{\rm exp}, \hat{E}} \approx \tau_{{\rm int}, \hat{E}} \propto L^{0.27}$, the exponent of which was estimated from the plots in Fig.~\ref{fig:e-3D}. We therefore estimate the dynamic critical exponent of the simple cubic lattice Ising model to be $z \approx 0.27$ in the worm update.

The exponential autocorrelation time in the present worm update is approximately 26 times as small as in the classical worm update, as shown in the inset of Fig.~\ref{fig:tau_exp-3D}, which is consistent with the asymptotic variances of the energy and the magnetic susceptibility estimators. Note that the summation of the autocorrelation function in the rescaled time is somewhat different from the rescaled integrated autocorrelation time~(\ref{tau_int_worm}) because of the existence of the constant $1/2$ in the definition~(\ref{tau_int}). Nonetheless, the asymptotic scaling is the same for the two quantities.

\begin{figure}  
\begin{center}
\includegraphics[width=\columnwidth]{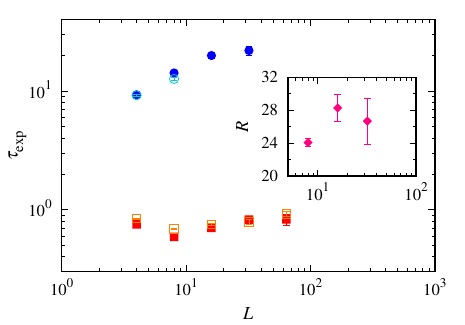}
\caption{(Color online) Exponential autocorrelation times of the energy (solid) and the magnetic susceptibility (open) estimators as a function of the system length in the classical (circles) and the present (squares) worm updates for the simple cubic lattice Ising model. The inset shows the ratio of the autocorrelation time of the energy estimator in the classical algorithm to the one in the present algorithm.}
\label{fig:tau_exp-3D}
\end{center}
\end{figure}

\section{Summary and discussion}
\label{sd}
We have proposed a modified worm algorithm for the Ising model and demonstrated performance improvement over the conventional worm algorithm at the critical temperature. The kinks of the present worm are located on bonds instead of sites of a lattice as shown in Figs.~\ref{fig:example} and~\ref{fig:ws}. The worm scattering probabilities are optimized using the directed worm framework and the geometric allocation approach as illustrated in Figs.~\ref{fig:go1} and~\ref{fig:go3}. We minimize the backscattering (rejection) probability and reduce it to zero in a wide range of temperatures, including the critical point. Moreover, we maximize the forward scattering probability to enhance further the diffusivity, or the diffusion constant, of the kink. Successful enhancement of the diffusivity is confirmed by observing the increased variance of the distribution of the distance between the two kinks, as displayed in Figs.~\ref{fig:P} and~\ref{fig:P_r}. As a result, the variance of the worm length, which is approximately proportional to the variance of the susceptibility estimator, is significantly reduced, as shown in Fig.~\ref{fig:sus-3D}\,(b).

We have discussed how to quantify the computational efficiency of the MCMC method and measure relevant quantities in the present approach. The relaxation rate is quantified by the exponential autocorrelation time, and the sampling efficiency is by the asymptotic variance, which is the prefactor of the asymptotic scaling of the statistical error squared.

The exponential autocorrelation times and the asymptotic variances in the present worm update are approximately only 4\% as large as in the classical (conventional) worm update for the simple cubic lattice Ising model as shown in Figs.~\ref{fig:e-3D},~\ref{fig:sus-3D}, ~\ref{fig:A}, and~\ref{fig:tau_exp-3D}. The present worm update is surprisingly even more efficient than the Wolff cluster update, although the exponent of the asymptotic variance is most likely the same. We expect the improvement over the classical algorithm to be attributed to the variance reduction of the worm length.

The dynamic critical exponent of the simple cubic lattice Ising model is estimated to be $ z \approx 0.27$ from fitting to a single power law $a L^z$, in which $a$ and $z$ are the fitting parameters. The resultant fit is statistically reasonable, producing a plausible mean square error of the regression $\chi^2/N_{\rm dof} \approx 1$, where $\chi^2$ is the sum of the squared residuals and $N_{\rm dof}$ is the number of degrees of freedom in the regression.

The estimate of the dynamic critical exponent is somewhat larger than the previous estimate: $z=\alpha / \nu \approx 0.174$, which was proposed in the Wolff cluster update\,\cite{CoddingtonB1992} and supported in the classical worm update\,\cite{DengGS2007}. This relation between the critical exponents was inferred from a numerical observation that the integrated autocorrelation time of the energy estimator is approximately proportional to the specific heat ($\propto L^{\alpha/\nu}$ asymptotically) at the critical temperature. We checked the ratio (not shown) of the autocorrelation time to the specific heat more precisely than the previous works did and found a slight but systematic increase as a function of $L$. This increase indicates $z > \alpha/\nu$, which is consistent with the direct fitting of the autocorrelation time. Note that although the total energy was measured in the extended state space in Ref.\,\citenum{DengGS2007}, the exponent of the autocorrelation time of the energy is expected to be the same for the original and the extended state space.

Our estimate $ z \approx 0.27$ is interestingly consistent with an estimate for the Wolff update, $z=0.24(2)$\,\cite{LiuPS2014}. This agreement suggests that the worm and the Wolff algorithms share the same exponent not only of the asymptotic variance but also of the exponential autocorrelation time.

We have estimated the exponents of the autocorrelation times: $L^{0.27} \propto \tau_{{\rm int}, \hat{E}} \approx \tau_{{\rm exp}, \hat{E}} \approx \tau_{{\rm exp}, \hat{\chi}} \gg \tau_{{\rm int}, \hat{\chi}} \sim L^{-0.7}$.

A lesson to learn from the present analysis is that we must be careful to estimate $\tau_{\rm exp}$ and needed thermalization (burn-in) steps. Because $\tau_{\rm int}$ is usually easier to estimate than $\tau_{\rm exp}$, in some (or probably many) cases, people roughly estimate $\tau_{\rm exp}$ assuming $\tau_{\rm exp} \sim \tau_{\rm int}$. This assumption is correct if the autocorrelation function is well approximated by a single exponential term and $\tau_{\rm exp} \gg 1$. If the autocorrelation function has more than one exponential terms, the integrated autocorrelation time is approximately given by $\tau_{\rm int} \sim c \, \tau_{\rm exp}$, where $c$ is the prefactor of the slowest decay. Therefore, $\tau_{\rm exp}$ can be much larger than $\tau_{\rm int}$ possibly in orders of magnitude as we have estimated $\tau_{{\rm int}, \hat{\chi}} \propto L^{-0.73}$ but $\tau_{{\rm exp}, \hat{\chi}} \propto L^{0.27}$ in the classical worm update. It is interesting that the prefactor decreases with the system length: $c \propto L^{-1.0}$.

The present approach can be generalized to a wide range of physical models to which the conventional worm algorithm has been applied, such as the $| \phi|^4$ model\,\cite{ProkofievS2001}, the Potts model\,\cite{MercadoEG2012}, the O($n$) loop model\,\cite{JankeNS2010,LiuDG2011,ShimadaJK2014}, and lattice QCD\,\cite{AdamsC2003}. The geometric allocation approach is expected to improve the computational efficiency of the directed worm update also for these systems. Our approach can be applied to frustrated models as well in combination with the dual worm formalism\,\cite{RakalaD2017}. In the meantime, an application of the lifting technique, which is another way to break the detailed balance, to the worm algorithm was recently proposed for the Ising model\,\cite{ElciGDNGD2018}. It is of interest to further combine our approach and the lifting technique. The performance of the present worm algorithm for other models needs to be investigated in the future.

\acknowledgments{The author is grateful to Synge Todo for the discussion on estimators in the worm algorithm. Some simulations were performed using computational resources of the Supercomputer Center at the Institute for Solid State Physics, the University of Tokyo. The author acknowledges support by KAKENHI under Grant No.\,16K17762 from JSPS.}

\bibliography{main}
\end{document}